\documentclass{iaus}
\usepackage{graphicx}

\title[Testing Planet Formation with Gaia] 
{Testing Planet Formation Models with Gaia $\mu$as Astrometry}

\author[A. Sozzetti et al.]   
{A. Sozzetti$^{1,2}$, S. Casertano$^3$, M.G. Lattanzi$^1$,
 A. Spagna$^1$, R. Morbidelli$^1$, R. Pannunzio$^1$, D. Pourbaix$^4$ 
 \and D. Queloz$^5$}

\affiliation{$^1$INAF- Astronomical Observatory of Torino, via Osservatorio 20,
I-10025, Pino Torinese (TO), Italy \\ email: {\tt sozzetti@oato.inaf.it}\\[\affilskip]
$^2$Harvard-Smithsonian Center for Astrophysics, 60 Garden Street, Cambridge, MA 02138, USA \\[\affilskip]
$^3$Space Telescope Science Institute, San Martin Drive, Baltimore, MD 21218, USA\\[\affilskip]
$^4$Institut d'Astronomie et d'Astrophysique, Universit\'e Libre de Bruxelles, CP. 226, Boulevard du Triomphe, 
1050 Bruxelles, Belgium\\[\affilskip]
$^5$Observatoire de Gen\`eve, 51 Ch. de Maillettes, 1290 Sauveny, Switzerland
}
\pubyear{2008}
\volume{xxx}  
\pagerange{???--???}
\jname{A Giant Step:from Milli- to Micro-arcsecond Astrometry}
\editors{A.C. Editor, B.D. Editor \& C.E. Editor, eds.}
\begin{document}

\maketitle

\begin{abstract}
In this paper, we first summarize the results of a large-scale double-blind tests campaign carried 
out for the realistic estimation of the Gaia potential in detecting and measuring 
planetary systems. Then, we put the identified capabilities in context by highlighting the 
unique contribution that the Gaia exoplanet discoveries will be able to bring to 
the science of extrasolar planets during the next decade.
\end{abstract}

\firstsection 
\section{Introduction}

Despite a few important successes (e.g., Bean et al. 2007, and references therein), 
astrometric measurements with mas precision have so far
proved of limited utility when employed as either a follow-up tool 
or to independently search for planetary mass companions orbiting 
nearby stars (see for example Sozzetti 2005, and references therein). 

In several past exploratory works (Casertano et al. 1996; 
Lattanzi et al. 1997, 2000; Sozzetti et al 2001, 2003), 
we have shown in some detail what space-borne astrometric observatories with 
$\mu$as-level precision, such as Gaia (Perryman et al. 2001), can achieve in terms of
search, detection and measurement of extrasolar planets of mass 
ranging from Jupiter-like to Earth-like. In those studies we adopted a qualitatively 
correct description of the measurements that each mission will carry out, and we estimated 
detection probabilities and orbital parameters using realistic, non-linear least squares fits to 
those measurements. 

Those exploratory studies, however, need updating and improvements. 
In the specific case of planet detection and measurement with Gaia, 
we have thus far largely neglected the
difficult problem of selecting adequate starting values for the
non-linear fits, using perturbed starting values instead.  The
study of multiple-planet systems, and in particular the determination
of whether the planets are coplanar---within suitable tolerances---is
incomplete.  The characteristics of Gaia have changed, in some ways
substantially, since our last work on the subject (Sozzetti et al 2003).  
Last but not least, in order to render the analysis truly independent 
from the simulations, these studies should be carried out in 
double-blind mode. 

We present here a substantial program of double-blind tests for 
planet detection with Gaia (preliminary findings were recently 
presented by Lattanzi et al. (2005)), with the three-fold goal of 
obtaining: a) an improved, more realistic assessment of the detectability
and measurability of single and multiple planets under a variety of
conditions, parametrized by the sensitivity of Gaia; 
b) an assessment of the impact of Gaia in critical areas of
planet research, in dependence on its expected capabilities; 
and c) the establishment of several Centers with a 
high level of readiness for the analysis of Gaia observations 
relevant to the study of exoplanets.

\section{Double-Blind Tests Campaign Results}

We carry out detailed simulations of Gaia observations of synthetic planetary systems 
and develop and utilize in double-blind mode independent software codes for the analysis of the data, 
including statistical tools for planet detection and different algorithms for single and multiple 
Keplerian orbit fitting that use no a priori knowledge of the true orbital parameters of the systems.

Overall, the results of our earlier works (e.g., Lattanzi et al. 2000; 
Sozzetti et al. 2001, 2003) are essentially confirmed, with the 
fundamental improvement due to the successful development of 
independent orbital fitting algorithms applicable to real-life data 
that do not utilize any a priori knowledge of the orbital parameters 
of the planets. In particular,  
the results of the T1 test (planet detection) indicate that planets down to 
astrometric signatures $\alpha\simeq 25$ $\mu$as, corresponding to 
$\sim 3$ times the assumed single-measurement error, can be detected 
reliably and consistently, with a very small number of false positives 
(depending on the specific choice of the threshold for detection). 
The results of the T2 test (single-planet orbital solutions) indicate that: 
1) orbital periods can be retrieved with very good accuracy (better than 
10\%) and small bias in the range $0.3\lesssim P\lesssim 6$ yrs, and 
in this period range the other orbital parameters and the planet 
mass are similarly well estimated. The quality of the solutions degrades 
quickly for periods longer than the mission duration, and in particularly 
the fitted value of $P$ is systematically underestimated; 
2) uncertainties in orbit parameters are well understood; 
3) nominal uncertainties obtained from the fitting procedure are a good
estimate of the actual errors in the orbit reconstruction.  Modest
discrepancies between estimated and actual errors arise only for planets
with extremely good signal (errors are overestimated) and for planets
with very long period (errors are underestimated); such discrepancies
are of interest mainly for a detailed numerical analysis, but they do
not touch significantly the assessment of Gaia's ability to find planets
and our preparedness for the analysis of perturbation data. 
The results of the T3 test (multiple-planet orbital solutions) 
indicate that 1) over 70\% of the simulated 
orbits under the conditions of the T3 test 
(for every two-planet system, periods shorter than 9 years and 
differing by at least a factor of two, $2\leq\alpha/\sigma_\psi\leq 50$, 
$e\leq 0.6$) are correctly identified; 
2) favorable orbital configurations (both planet with 
periods $\leq 4$ yr and astrometric signal-to-noise ratio $\alpha/\sigma_\psi\geq 10$, 
redundancy of over a factor of 2 in the number of observations) 
have periods measured to better than 10\% accuracy $> 90\%$ 
of the time, and comparable results hold for other orbital 
elements; 3) for these favorable cases, only a modest 
degradation of up to $10\%$ in the fraction of well-measured orbits 
is observed with respect to single-planet solutions with comparable 
properties; 4) the overall results are mostly insensitive to the relative 
inclination of pairs of planetary orbits; 5) over 80\% of the favorable 
configurations have $i_\mathrm{rel}$ measured to better 
than 10 degrees accuracy, with only mild dependencies on its 
actual value, or on the inclination angle with respect to 
the line of sight of the planets; 
6) error estimates are generally accurate, particularly for 
fitted parameters, while modest discrepancies (errors are 
systematically underestimated) arise between 
formal and actual errors on $i_\mathrm{rel}$.

\begin{figure}
\centering
\includegraphics[width=0.75\textwidth]{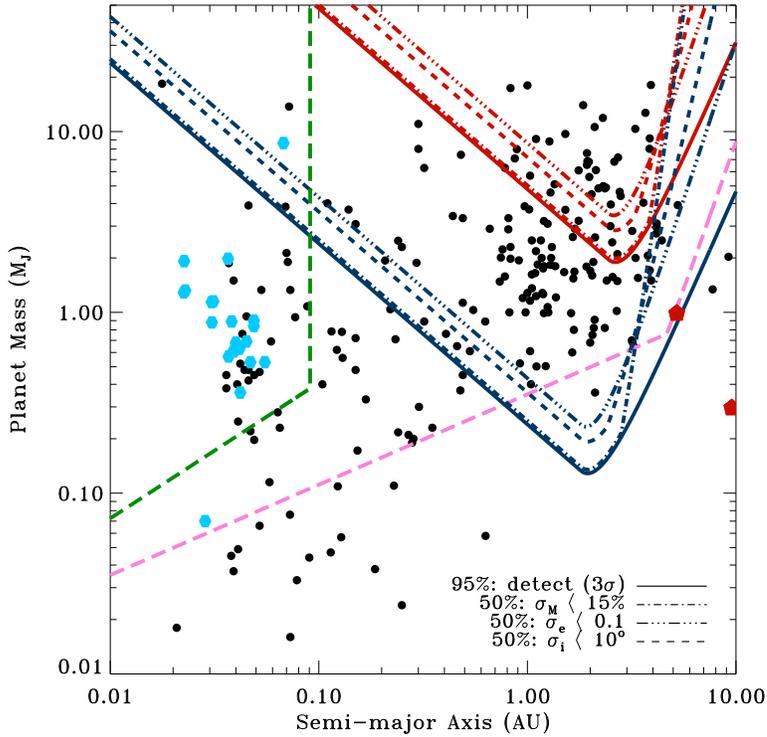}
\caption{Gaia discovery space for planets of given mass and orbital radius compared 
to the present-day sensitivity of other indirect detection methods, namely Doppler 
spectroscopy and transit photometry. Red curves of different styles 
(for completeness in planet detection and orbit measurement to given accuracy) 
assume a 1-$M_\odot$ G dwarf primary at 200 pc, 
while the blue curves are for a 0.5-$M_\odot$ M dwarf at 25 pc. The radial velocity 
curve (pink line) is for detection at the $3\times\sigma_\mathrm{RV}$ level, assuming 
$\sigma_\mathrm{RV} = 3$ m s$^{-1}$, $M_\star = 1 M_\odot$, and 10-yr survey duration. 
For transit photometry (green curve), 
$\sigma_V = 5$ milli-mag, $S/N = 9$, $M_\star=1$ $M_\odot$, $R_\star = 1$ $R_\odot$, 
uniform and dense ($> 1000$ datapoints) sampling. 
Black dots indicate the inventory of exoplanets as of October 2007. Transiting systems 
are shown as light-blue filled pentagons. Jupiter and Saturn are also shown as red pentagons.}
\label{detmeas}
\end{figure}

\section{The Gaia Legacy}

\begin{table}[tbh]
\begin{minipage}{0.5\linewidth}
\begin{center}
   \caption{Number of giant planets that could be detected and measured by Gaia, as a 
   function of increasing distance. Starcounts are obtained using the Besancon model of 
   stellar population synthesis (Bienaym\'e et al. 1987), while the Tabachnik \& Tremaine (2002) model 
for estimating planet frequency as a function of mass and orbital period is utilized.}
   \renewcommand{\arraystretch}{1.4}
   \setlength\tabcolsep{7pt}
{\tiny
      \begin{tabular}{|c|c|c|c|c|c|c|}
       \hline\noalign{\smallskip}
       $\Delta d$ & $N_\star$  & $\Delta a$ & $\Delta M_p$ &  $N_{\rm d}$ &  $N_{\rm m}$ \\
       (pc) & & (AU) & ($M_J$) & & \\
              \noalign{\smallskip}
       \hline
       \noalign{\smallskip}
0-50 & $1\times10^4$ & 1.0 - 4.0 & 1.0 - 13.0 & $1400$ & $ 700$\\ \hline 
50-100 & $5\times10^4$ & 1.0 - 4.0 & 1.5 - 13.0 & $2500$ & $ 1750$\\ \hline 
100-150 & $1\times10^5$ & 1.5 - 3.8 & 2.0 - 13.0& $2600$ & $ 1300$\\\hline 
150-200 & $3\times10^5$ & 1.4 - 3.4 & 3.0 - 13.0& $2150$ & $ 1050$\\\hline 

   \end{tabular}
}
\label{nplan}
\end{center}
\end{minipage}
\hspace{0.5cm}
\begin{minipage}{0.5\linewidth}
\begin{center}
   \caption{Number of multiple-planet systems that Gaia could potentially detect, measure, 
   and for which coplanarity tests could be carried out successfully.}
   \renewcommand{\arraystretch}{1.4}
   \setlength\tabcolsep{7pt}
{\tiny
      \begin{tabular}{|l|c|}
       \hline\noalign{\smallskip}
        Case & Number of Systems \\
              \noalign{\smallskip}
       \hline
       \noalign{\smallskip}
Detection & $\sim 1000$\\ \hline
Orbits and masses to  & \\ 
better than 15-20\% accuracy & $\sim 400-500$ \\ \hline
Successfull  & \\
coplanarity tests & $\sim 150$\\ \hline
   \end{tabular}
}
\label{nmult}
\end{center}
\end{minipage}
\end{table}

In Figure~\ref{detmeas} we show Gaia's discovery space in terms of detectable and measurable 
planets of given mass and orbital separation around stars of given mass at a given distance 
from Earth (see caption for details). From the Figure, one would then conclude that Gaia 
could discover and measure massive giant planets ($M_p \gtrsim 2-3$ $M_\mathrm{J}$) 
with $1<a<4$ AU orbiting solar-type stars as far as the nearest star-forming regions, 
as well as explore the domain of Saturn-mass planets with similar 
orbital semi-major axes around late-type stars within 30-40 pc. These results can be turned 
into a number of planets detected and measured by Gaia, using Galaxy models and the current 
knowledge of exoplanet frequencies. By inspection of Tables~\ref{nplan} and~\ref{nmult}, 
we then find that Gaia could measure accurately thousands of giant planets, and accurately 
determine coplanarity (or not) for a few hundred multiple systems with favorable configurations. 

In conclusion, Gaia's main strength continues to be the ability 
to measure actual masses and orbital parameters for possibly 
thousands of planetary systems. The Gaia data have the potential to 
a) significantly refine our understanding of the statistical properties 
of extrasolar planets: the predicted database of several 
thousand extrasolar planets with well-measured properties will allow 
for example to test the fine structure of giant planet parameters 
distributions and frequencies, and to investigate their possible changes 
as a function of stellar mass with unprecedented resolution; 
b) help crucially test theoretical models of 
gas giant planet formation and migration: for example, specific predictions 
on formation time-scales and the role of varying metal content in the 
protoplanetary disk will be probed with unprecedented statistics thanks 
to the thousands of metal-poor stars and hundreds of young stars 
screened for giant planets out to a few AUs ; c) improve our comprehension of the 
role of dynamical interactions in the early as well as long-term 
evolution of planetary systems: for example, the measurement of orbital parameters 
for hundreds of multiple-planet systems, including meaningful coplanarity 
tests will allow to discriminate between various proposed mechanisms 
for eccentricity excitation; d) aid in the understanding of 
direct detections of giant extrasolar planets: for example, actual mass estimates and full orbital 
geometry determination for suitable systems will 
inform direct imaging surveys about where and when to point, in order to 
estimate optimal visibility, and will help in the modeling and interpretation 
of giant planets' phase functions and light curves;  
e) provide important supplementary data for the optimization of the 
target selection for Darwin/TPF: for example, all F-G-K-M stars within the useful volume ($\sim 25$ pc) 
will be screened for Jupiter- and Saturn-sized planets out to several AUs, and 
these data will help probing the long-term dynamical stability of their 
Habitable Zones, where terrestrial planets may have formed, and maybe found.

\end{document}